\documentclass[10pt,aps,prl,twocolumn,superscriptaddress,showpacs,preprintnumbers,amsmath,amssymb]{revtex4-1}

\usepackage{amsmath}
\usepackage{graphicx}
\usepackage{dcolumn}
\usepackage{bm}
\usepackage{color}

\begin{document}

\title{Estimating topological properties of weighted networks from limited information} 

\author{Giulio Cimini}
\email{giulio.cimini@roma1.infn.it}
\affiliation{Istituto dei Sistemi Complessi (ISC-CNR) UoS ``Sapienza'' University of Rome, P.le A. Moro 5, 00185 Rome (Italy)}
\author{Tiziano Squartini}
\email{tiziano.squartini@roma1.infn.it}
\affiliation{Istituto dei Sistemi Complessi (ISC-CNR) UoS ``Sapienza'' University of Rome, P.le A. Moro 5, 00185 Rome (Italy)}
\author{Andrea Gabrielli}
\affiliation{Istituto dei Sistemi Complessi (ISC-CNR) UoS ``Sapienza'' University of Rome, P.le A. Moro 5, 00185 Rome (Italy)}
\affiliation{IMT Institute for Advanced Studies, Piazza San Ponziano 6, 55100 Lucca (Italy)}
\author{Diego Garlaschelli}
\affiliation{Lorentz Institute for Theoretical Physics, University of Leiden, Niels Bohrweg 2, 9506 Leiden (Netherlands)}
\date{\today}

\begin{abstract}
A fundamental problem in studying and modeling economic and financial systems is represented by privacy issues, which put severe limitations on the amount of accessible information. 
Here we introduce a novel, highly nontrivial method to reconstruct the structural properties of complex weighted networks of this kind using only partial information: 
the total number of nodes and links, and the values of the strength for all nodes. The latter are used as {\em fitness} to estimate the unknown node degrees through a standard {\em configuration model}. 
Then, these estimated degrees and the strengths are used to calibrate an {\em enhanced configuration model} in order to generate ensembles of networks intended to represent the real system. 
The method, which is tested on real economic and financial networks, while drastically reducing the amount of information needed to infer network properties, 
turns out to be remarkably effective---thus representing a valuable tool for gaining insights on privacy-protected socioeconomic systems.
\end{abstract}

\maketitle 



Reconstructing the statistical properties of a network when only partial information is available represents a key unsolved problem 
in the field of statistical physics of complex systems \cite{Clauset2008,Mastromatteo2012}. Yet, addressing this issue can bring to many concrete applications.
A paramount example is provided by financial networks, where nodes represent financial institutions and edges stand for the various types of financial ties---such as loans 
or derivative contracts. These ties result in dependencies among institutions and constitute the ground for the propagation of financial distress across the network. 
However, due to confidentiality issues, the information that regulators are able to collect on mutual exposures is very limited \cite{Wells2004}, and this hinders the analysis 
of the system resilience to the default or distress on one or more institutions---which depends on the structure of the whole network \cite{Battiston2012a,Battiston2012b}. 
Typically, the analysis of systemic risk has been pursued by trying to reconstruct the unknown links of the network using Maximum Entropy algorithms 
\cite{Lelyveld2006,Degryse2007,Mistrulli2011}. These approaches, also known as ``dense reconstruction'' methods, assume that the network is fully connected 
and estimate link weights via a maximum homogeneity principle, looking for the weighted adjacency matrix with minimal distance from the uniform matrix 
that also satisfies the imposed constraints---represented for instance by the budget of individual banks. The strongest limitation of these algorithms lies in the hypothesis that the network 
is fully connected. In fact, not only empirical networks show a very heterogeneous distribution of the connectivity, but such dense reconstruction was shown 
to lead to systemic risk underestimation \cite{Mastromatteo2012,Mistrulli2011}. More refined methods like ``sparse reconstruction'' algorithms \cite{Mastromatteo2012} 
allow to obtain a matrix with an arbitrary level of heterogeneity, but still leave open the question of what value of heterogeneity would be appropriate; 
moreover, even when link density is correctly recovered, systemic risk is again underestimated because of the homogeneity principle used to build the network. 
A more recent approach \cite{Musmeci2013,Caldarelli2013} instead uses the limited topological information available on the network to generate 
an ensemble of {\em exponential random graphs} (ERG) through the {\em configuration model} (CM) \cite{Park2004}, where the Lagrange multipliers defining it are replaced 
by {\em fitnesses} \cite{Caldarelli2002}, \emph{i.e.}, node-specific properties assumed to be known---in a way similar to fitness-dependent network models \cite{Garlaschelli2004}. 
The estimation of network properties is then carried out within such fitness-induced ensemble. This method overcomes the limitations of its predecessors, 
but it still suffers from the drawback of being applicable only to binary networks---whereas, the analysis of systemic risk is generally carried out within the weighted representation of the networked system. 

Here we aim at overcoming all the limitations of these methods and build an innovative and effective procedure to reconstruct weighted networks, resorting on a minimal amount of available information: 
the total number of connections and the values of the strength for each node---which will play the role of node fitness. 
In a nutshell, our method consists in estimating the number of connections for each node via the standard CM calibrated on the fitnesses, 
and then in using these values as well as node strengths to assess individual link weights through an {\em enhanced configuration model} (ECM) \cite{Mastrandrea2014}. 
To validate our method, we use two real instances of economic and financial systems. The first one is the World Trade Web (WTW) \cite{Gleditsch2002}, {\em i.e.}, 
the network whose $N$ nodes represent countries and whose $L$ links represent trade volumes---so that the weight $w_{ij}$ of the link between nodes $i$ and $j$ 
is the total monetary flux between these countries resulting from the import/export among them \cite{foot1a}. The second one is the (E-mid) interbank money market \cite{DeMasi2006}, 
where now the nodes represent banks and $w_{ij}$ is the total amount of loans ({\em i.e.}, of liquidity exchanged) between banks $i$ and $j$ \cite{foot1b}. 
In both cases, the strength of node $i$ is defined as $s_i^*=\sum_j w_{ij}$, while its degree or number of partners is $k_i^*=\sum_j a_{ij}$ 
(where $a_{ij}:=\lim_{\varepsilon\rightarrow0}[1+\varepsilon/w_{ij}]^{-1}$). Since we have full information on these networks, we will be able to assess unambiguously 
the accuracy of our method in estimating their topological properties. 



Our network reconstruction procedure builds on two complementary network generation models.
The CM \cite{Park2004}, a particular class of ERG model \cite{Dorogovtsev2010}, consists in generating an ensemble $\Omega_{CM}$ of networks which is maximally random---except 
for the ensemble average of the node degrees $\{\langle k_i \rangle_{\Omega_{CM}}\}_{i=1}^N$ that are constrained to the observed values $\{k_i^*\}_{i=1}^N$. 
The probability distribution over $\Omega_{CM}$ is defined via a set of Lagrange multipliers $\{x_i\}_{i=1}^N$ (one for each node), whose values can be set 
to satisfy the equivalence $\langle k_i \rangle_{\Omega_{CM}} \equiv k_i^*$ $\forall i$ \cite{Squartini2011}. 
The ensemble probability that any two nodes $i$ and $j$ are connected is given by: 
\begin{equation}
 p_{ij}=\frac{x_ix_j}{1+x_ix_j},
\label{eq:prob linking}
\end{equation}
so that $x_i$ quantifies the ability of node $i$ to create links with other nodes. 
The ECM \cite{Mastrandrea2014} is instead obtained by specifying both the mean degree and strength sequences $\{k_i^*\}_{i=1}^N$ and $\{s_i^*\}_{i=1}^N$. 
In this case, two Lagrange multipliers $\{a_i,b_i\}$ are associated to each node $i$, so that the ensemble probability $q_{ij}$ that any two nodes $i$ and $j$ are connected 
and the ensemble average $\langle w_{ij} \rangle$ for the weight of such link become \cite{Garlaschelli2009}:
\begin{equation}
 q_{ij}=\frac{a_ia_jb_ib_j}{1+a_ia_jb_ib_j-b_ib_j},\qquad \langle w_{ij} \rangle=\frac{q_{ij}}{1-b_ib_j}.
\label{eq:prob linking 2}
\end{equation}
On the other hand, the {\em fitness} model \cite{Caldarelli2002} assumes the network topology to be determined by an intrinsic property (fitness) 
associated with each node. This approach has been successfully used in the past to model several economic networks, 
including the network of equity investments in the stock market \cite{Garlaschelli2005}, the E-mid \cite{DeMasi2006} and the WTW~\cite{Garlaschelli2004}. 
Note that fitnesses are often used within the ERG framework provided an assumed connection between them 
and the Lagrange multipliers. Our method builds exactly on such assumption. 

\begin{figure}[t!]
\begin{center}
\includegraphics[width=8.6cm]{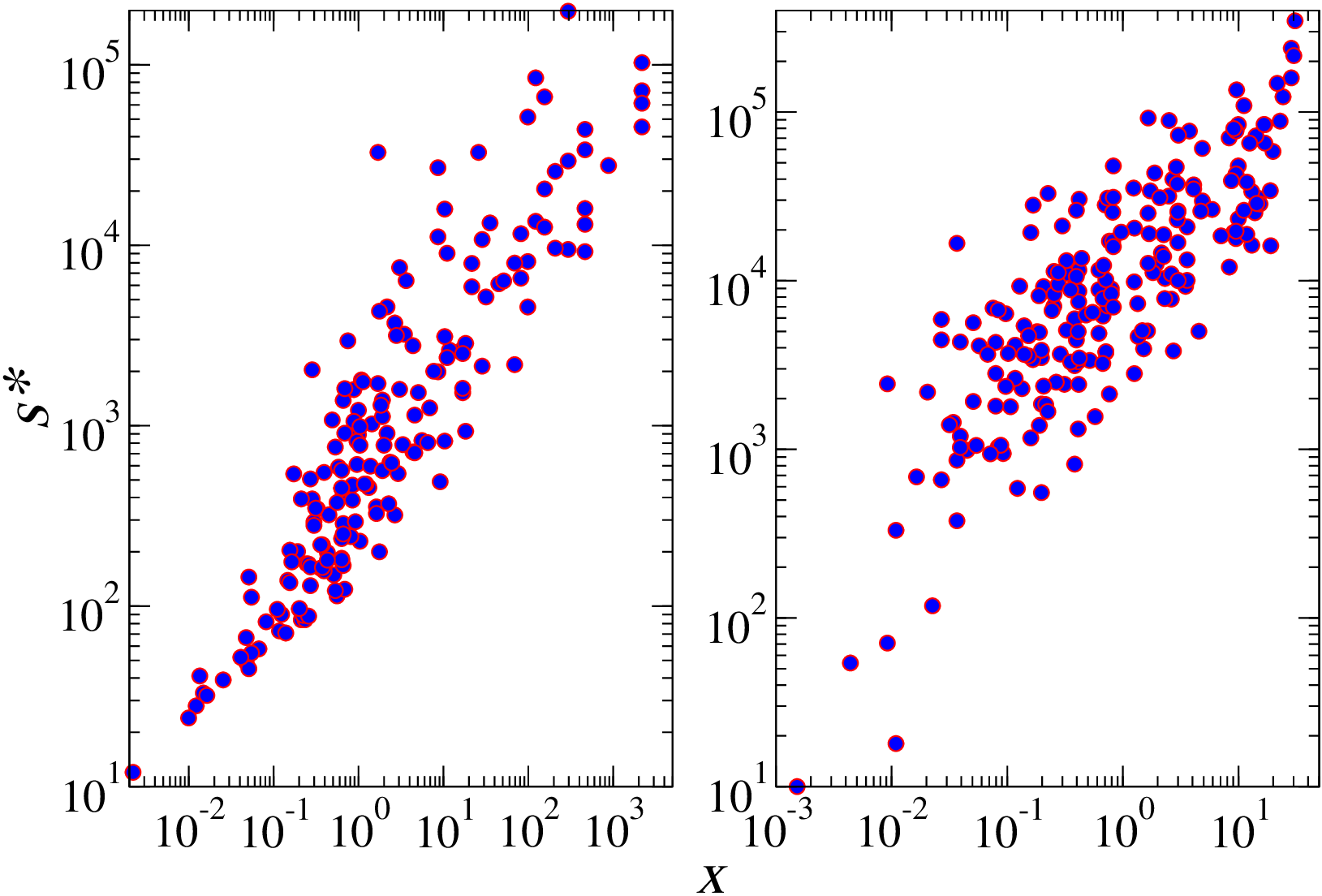}
\end{center}
\vspace{-0.5cm}
\caption{Relation between node strengths $\{s^*\}$ and their degree-induced Lagrange multipliers $\{x\}$ from CM (obtained by knowing the whole degree sequence). 
The linearity of such relation is at the basis of assumption 2 of our method that $x_i\propto s_i^*$ $\forall i$ accurately describes the binary network topology. 
Left panel refers to WTW, right panel to E-mid.}
\label{fig:corr}
\end{figure}

Given these ingredients, we now formulate the statistical procedure at the basis of our method. 
We aim at finding the most probable estimate for $X(G_0)$, {\em i.e.}, the value  of a topological property $X$ 
for the real network $G_0$ that we want to reconstruct. Such estimate has to rely on and be compatible with some constraints, given by the incomplete information we have on $G_0$: 
the total number of nodes $N$ and links $L$, and the whole strength sequence $\{s_i^*\}_{i=1}^N$. We build on two important assumptions:
\begin{enumerate}
 \item $G_0$ can be seen as drawn from an appropriate ECM ensemble $\Omega_{ECM}$, so that $X(G_0)$ can be estimated as $\langle X \rangle_{\Omega_{ECM}}$;
 \item The strengths $\{s_i^*\}_{i=1}^N$ represent degree-induced node fitnesses, and are thus assumed to be proportional to the Lagrange multipliers $\{x_i\}_{i=1}^N$ of the CM 
 via a universal parameter $z$: $x_i\equiv\sqrt{z}s_i^*$ $\forall i$.
\end{enumerate}
The first assumption allows us to map the problem of evaluating $X(G_0)$ into that of choosing the optimal ECM ensemble $\Omega_{ECM}$ compatible with the known constraints on $G_0$. 
In other words, the question to address becomes: what ECM ensemble is the most appropriate to extract the real network $G_0$ from, given that we  know only partial information? 
Then, once $\Omega_{ECM}$ is determined, we can use the average $\langle X \rangle_{\Omega_{ECM}}$ as a good estimation for $X(G_0)$. 
However, in order to build an appropriate $\Omega_{ECM}$, we need to know not only the strengths but also the degrees for all the nodes, and this is where assumption 2 comes in handy: 
the unknown degrees can be estimated within a CM ensemble $\Omega_{CM}$ built using the strengths as degree-induced fitnesses (Figure \ref{fig:corr}). 

Technically, our method consists in the following operative steps. 
I) We first find the unknown parameter $z$ that defines $\Omega_{CM}$ (see assumption 2) 
by comparing the average number of links of a network belonging to $\Omega_{CM}$ with the (known) total number $L(G_0)$ of links in $G_0$: 
\begin{equation}
\langle L \rangle_{\Omega_{CM}} = \frac{1}{2} \sum_i \sum_{j (\neq i)} \frac{zs_i^*s_j^*}{1+zs_i^*s_j^*} \equiv L(G_0)
\label{eq:L} 
\end{equation} 
(since $\{s_i^*\}_{i=1}^N$ are known, eq.~(\ref{eq:L}) is an algebraic equation in $z$). 
We then use this $z$ to estimate the unknown degrees through eq.~(\ref{eq:prob linking}): 
\begin{equation}
 \langle k_i\rangle_{\Omega_{CM}} = \sum_{j (\neq i)} p_{ij} = \sum_{j (\neq i)} \frac{zs_i^*s_j^*}{1+zs_i^*s_j^*}\qquad\forall i.
\label{eq:degrees}
\end{equation}
II) We use the degrees estimated in this way in the system of $2N$ nonlinear equations that define the ECM:
\begin{equation}
\label{eq:ecm}
 \left\{ \begin{aligned}
        \langle k_i\rangle_{\Omega_{CM}} = & \sum_{j(\neq i)}\frac{a_ia_jb_ib_j}{1+a_ia_jb_ib_j-b_ib_j}\\
	s_i^*(G_0) = & \sum_{j(\neq i)}\frac{a_ia_jb_ib_j}{(1+a_ia_jb_ib_j-b_ib_j)(1-b_ib_j)}
	\end{aligned}
 \right. \forall i.
\end{equation}
The solution is the set of Lagrange multipliers $\{a_i,b_i\}_{i=1}^N$ that define the ECM ensemble---through the linking probabilities $\{q_{ij}\}_{i,j=1}^N$ 
and the average weights $\{\langle w_{ij}\rangle\}_{i,j=1}^N$ as of eq.~(\ref{eq:prob linking 2})---and allow to compute $\langle X \rangle_{\Omega_{ECM}}$, either analytically or numerically. 



\begin{figure}[t!]
\begin{center}
\includegraphics[width=8.6cm]{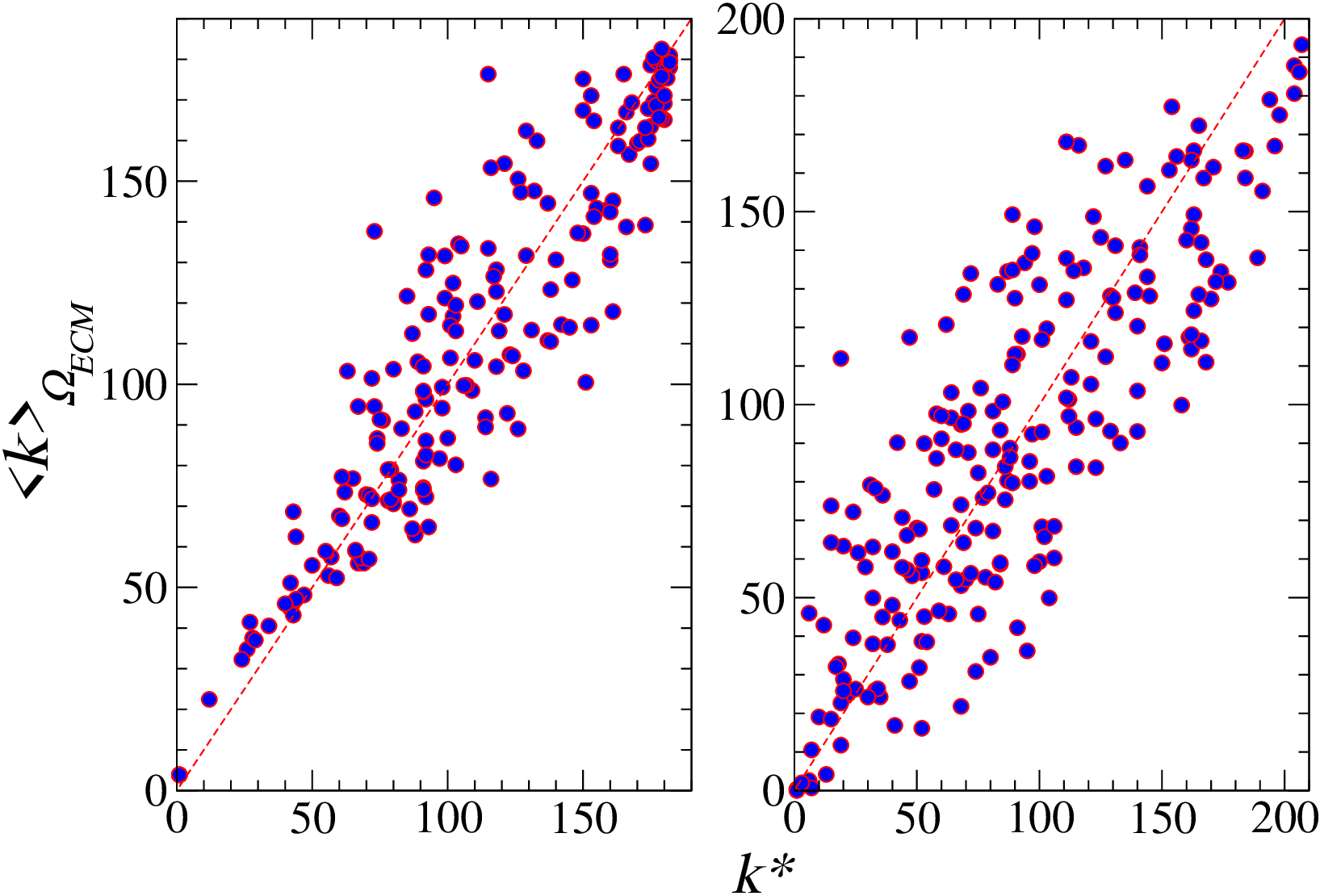}
\end{center}
\vspace{-0.5cm}
\caption{Relation between $k^*$ and $\langle k\rangle_{\Omega_{ECM}}$ for WTW (left panel) and E-mid (right panel).}
\label{fig:base_test}
\end{figure}

\begin{figure*}[t!]
\begin{center}
\includegraphics[width=16cm]{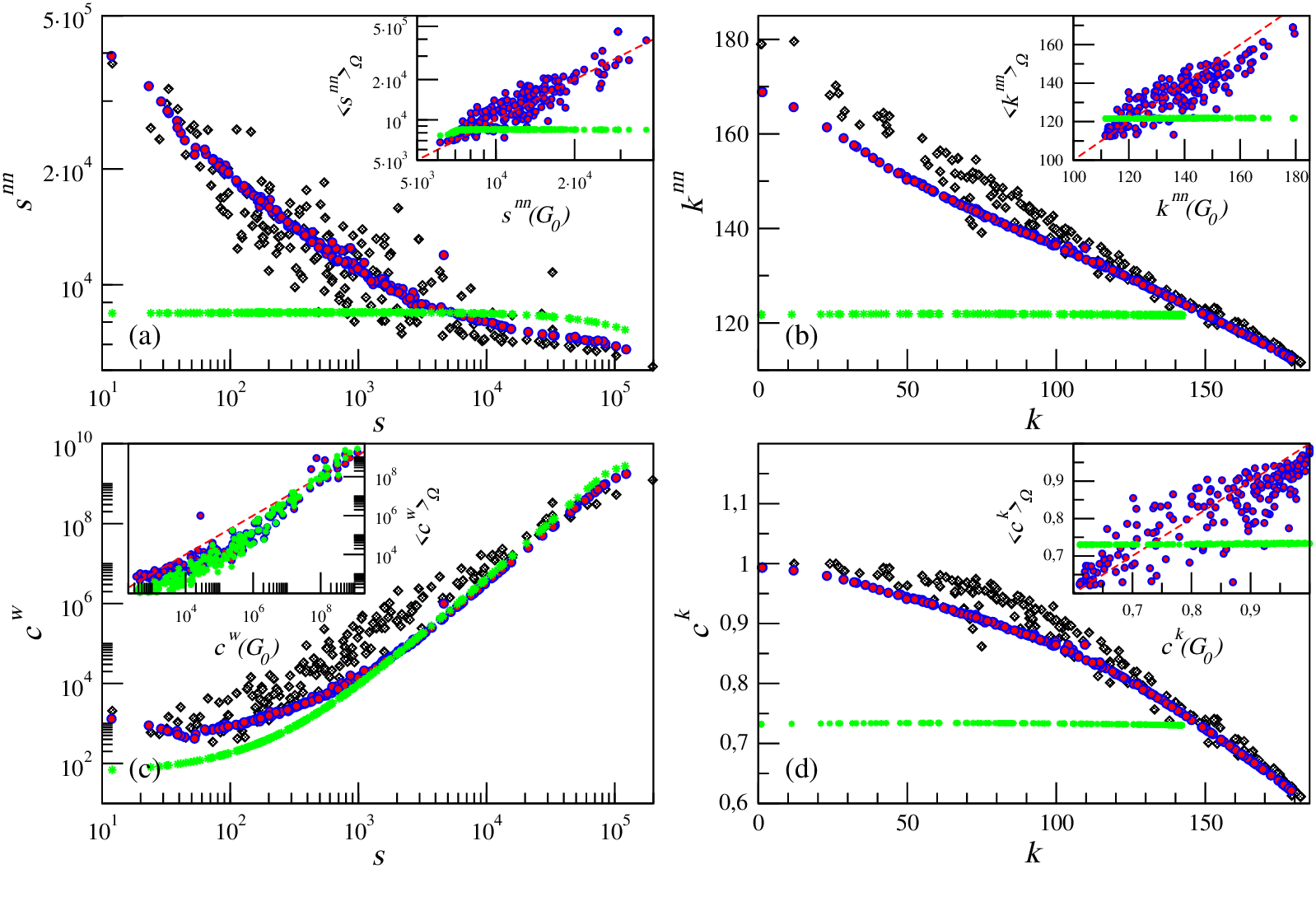}
\includegraphics[width=16cm]{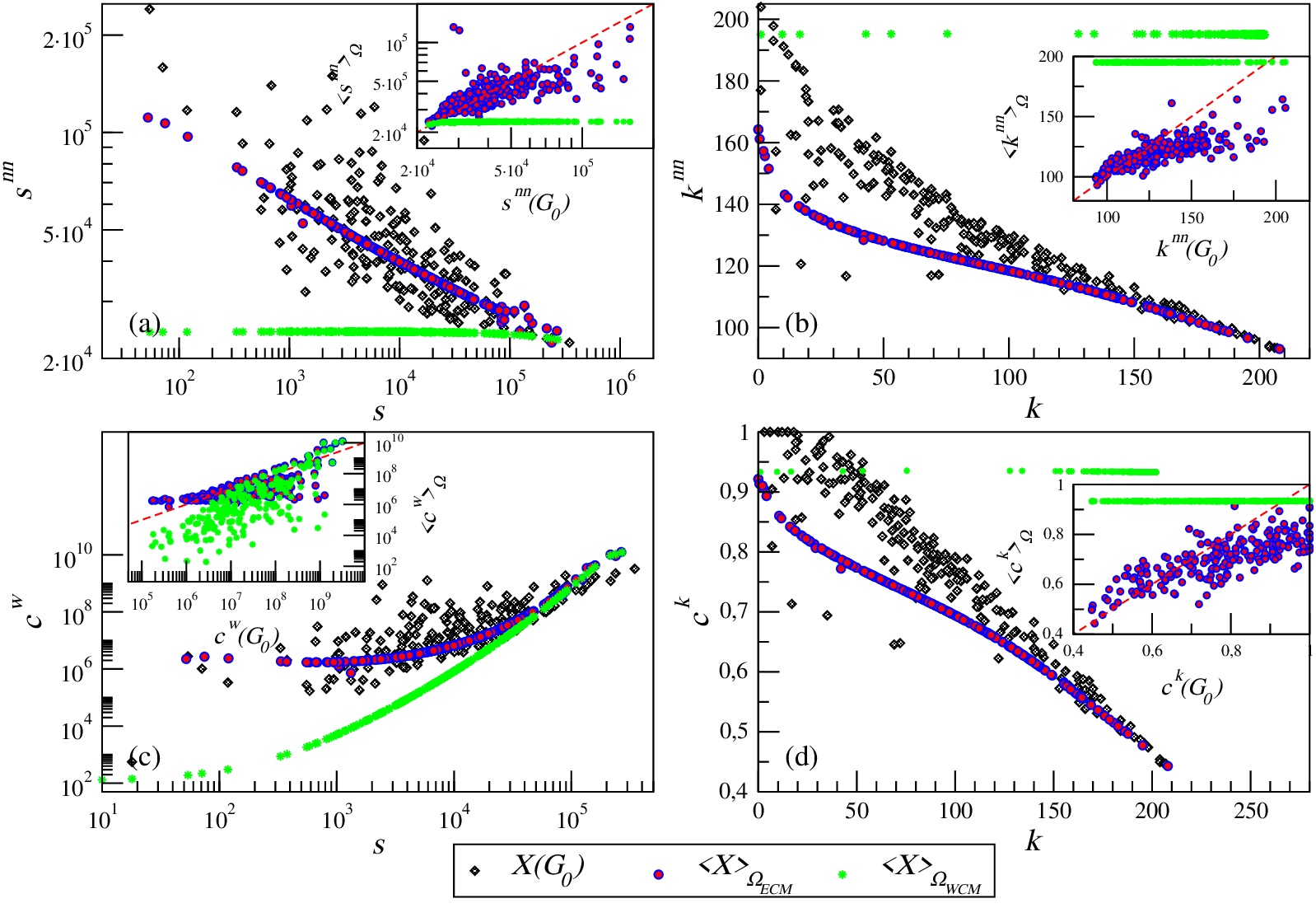}
\end{center}
\vspace{-0.5cm}
\caption{Scatter plots of $s$ vs $s^{nn}$ (a), $k$ vs $k^{nn}$ (b), $s$ vs $c^w$ (c) and $k$ vs $c^k$ (d) 
for the real quantities ($X(G_0)$), those estimated by our method ($\langle X\rangle_{\Omega_{ECM}}$), and those computed by a WCM-based reconstruction ($\langle X\rangle_{\Omega_{WCM}}$). 
Insets: relations $X(G_0)$ vs $\langle X\rangle_{\Omega_{ECM}}$ and $X(G_0)$ vs $\langle X\rangle_{\Omega_{WCM}}$ for the same quantities. Upper plots refer to WTW, lower plots to E-mid.} 
\label{fig:next_test}
\end{figure*}

In order to check whether $\Omega_{ECM}$ defined above is a proper ensemble to draw the real network $G_0$ from, 
we first compare for each node $i$ the degree $k_i^*(G_0)$ of the real network and $\langle k_i\rangle_{\Omega_{ECM}}=\sum_{j(\neq i)}q_{ij}$ estimated through our method \cite{foot3}. 
As Figure \ref{fig:base_test} shows, this results in a scattered cloud around the identity, whose behavior reflects the noisy yet very high correlation between strengths and degrees---as 
we are not using the real $k^*(G_0)$ in eq.~(\ref{eq:ecm}) but $\langle k_i\rangle_{\Omega_{CM}}$ obtained from the CM induced by node strengths. 
We move further and focus on the topological properties that are commonly regarded as the most significant for describing a weighted network structure: 
the {\em average nearest neighbors strength} 
\begin{equation}\label{eq.s_nn}
s^{nn}_i:=\frac{\sum_{j(\neq i)}a_{ij}s_{j}}{k_i}=\frac{\sum_{j(\neq i)}\sum_{k(\neq i,j)}a_{ij}w_{jk}}{\sum_{j(\neq i)}a_{ij}}
\end{equation}
and the {\em weighted clustering coefficient} 
\begin{equation}\label{eq.clu_w}
c^w_i:=\frac{\sum_{j(\neq i)}\sum_{k(\neq i,j)}w_{ij}w_{ik}w_{jk}}{\sum_{j(\neq i)}\sum_{k(\neq i,j)}a_{ij}a_{ik}},
\end{equation}
together with the binary version of these quantities: the {\em average nearest neighbors degree} 
\begin{equation}\label{eq.k_nn}
k^{nn}_i:=\frac{\sum_{j(\neq i)}a_{ij}k_{j}}{k_i}=\frac{\sum_{j(\neq i)}\sum_{k(\neq i,j)}a_{ij}a_{jk}}{\sum_{j(\neq i)}a_{ij}}
\end{equation}
and the {\em binary clustering coefficient} 
\begin{equation}\label{eq.clu_k}
c^k_i:=\frac{\sum_{j(\neq i)}\sum_{k(\neq i,j)}a_{ij}a_{ik}a_{jk}}{\sum_{j(\neq i)}\sum_{k(\neq i,j)}a_{ij}a_{ik}}.
\end{equation}
The ECM ensemble averages for these quantities are obtained from eqn.~(\ref{eq.s_nn}-\ref{eq.clu_k}) by replacing the binary adjacency matrix elements 
$a_{ij}$ with the linking probabilities $q_{ij}$, and the real link weights $w_{ij}$ with their ensemble averages $\langle w_{ij}\rangle$. 
Figure \ref{fig:next_test} shows a remarkable agreement between the values of these quantities computed on $G_0$ and their ECM 
ensemble averages---which can therefore be used as good estimates for the real quantities $X(G_0)$. 
Such test reveals the effectiveness of our method in reconstructing the topological properties of the real network.



It is important to remark that the applicability of our method strongly depends on the accuracy of assumption 2, \emph{i.e.} on whether the CM induced by node strengths is able to provide good estimates 
for the unknown degrees. This is indeed the case of the WTW \cite{Garlaschelli2004} and the E-mid \cite{DeMasi2006}, but also of other economic and financial networks of different nature \cite{Garlaschelli2005}.
Another important remark is that our method is based on a combination of CM and ECM rather than directly on the Weighted Configuration Model (WCM) \cite{Squartini2011}, 
because the latter not only fails to reproduce the network topological properties (as shown by Figure \ref{fig:next_test}), but also predicts a far denser network than observed. 
This happens not because strengths carry a ``lower level'' information than that of degrees---rather, they can be used to infer the degrees themselves, and this is what our method points out: 
the information on strength values should not be used to directly reconstruct the network, but to estimate the degree first, and only then to compute the quantities of interest.
In this respect, note that using directly the knowledge of the strength sequence and number of links as fixed constraints to build a maximum-entropy ensemble 
would result in a different mathematical expressions. In particular, we would arrive at a variant of eq.~(\ref{eq:prob linking 2}) where $a_i=a$ $\forall i$. We have checked that, just like the WCM, 
this model gives a bad prediction of the network, leading to the conclusion that inferring the expected degrees first through eq.~(\ref{eq:degrees}) is a crucial step of the approach we are using here: 
the information on links presence is indispensable to achieve a faithful network reconstruction. 

Further work is needed to address several issues that remain open, including testing the accuracy of our method in estimating higher-order topological properties. 
Possibly, for these cases the method could require a larger initial information to obtain the same effectiveness. 
Nevertheless, in its present version our method exploits a very limited information, which is indeed minimal but also often available for economic and financial systems: 
besides global statistics ($N$ and $L$), the strengths (that can be the operating revenue of firms, or the tier-1 capital of banks) are or should be accessible public data. 
In conclusion, our method is particularly useful to overcome the lack of topological information that often hampers systemic risk estimation in financial networks. 
More generally, our method can be applied to any network representing a set of dependencies among components in a complex system for which the available information is limited, 
and it is thus of general interest in the field of statistical physics of networks.
\newline

This work was supported by the EU project GROWTHCOM (611272), the Italian PNR project CRISIS-Lab, the EU project MULTIPLEX (317532) and the Netherlands Organization for Scientific Research (NWO/OCW).
DG acknowledges support from the Dutch Econophysics Foundation (Stichting Econophysics, Leiden, the Netherlands) with funds from beneficiaries of Duyfken Trading Knowledge BV (Amsterdam, the Netherlands).


\begin{thebibliography}{23}
\bibitem{Clauset2008} A. Clauset, C. Moore, M. E. J. Newman. Nature {\bf 453}(7191), 98-101 (2008).
\bibitem{Mastromatteo2012} I. Mastromatteo, E. Zarinelli, M. Marsili. J. Stat. Mech. {\bf 2012}(03), P03011 (2012).
\bibitem{Wells2004} S. Wells, \emph{Financial interlinkages in the United Kingdom's interbank market and the risk of contagion} (Bank of England's Working paper 230, 2004). 
\bibitem{Battiston2012a} S. Battiston, D. Gatti, M. Gallegati, B. Greenwald, J. Stiglitz. J. Econ. Dyn. Contr. {\bf 36}(8), 1121-1141 (2012).
\bibitem{Battiston2012b} S. Battiston, M. Puliga, R. Kaushik, P. Tasca, G. Caldarelli. Sci. Rep. {\bf 2}, 541 (2012).
\bibitem{Lelyveld2006} I. van Lelyveld, F. Liedorp. Int. J. Cent. Bank. {\bf 2}, 99-134 (2006).
\bibitem{Degryse2007} H. Degryse, G. Nguyen. Int. J. Cent. Bank. {\bf 3}, 123-171 (2007).
\bibitem{Mistrulli2011} P. Mistrulli. J. Bank. Fin. {\bf 35}(5), 1114-1127 (2011).
\bibitem{Musmeci2013} N. Musmeci, S. Battiston, G. Caldarelli, M. Puliga, A. Gabrielli. J. Stat. Phys. {\bf 151}(3-4), 720-734 (2013).
\bibitem{Caldarelli2013} G. Caldarelli, A. Chessa, A. Gabrielli, F. Pammolli, M. Puliga. Nat. Phys. {\bf 9}, 125 (2013).
\bibitem{Park2004} J. Park, M. E. J. Newman. Phys. Rev. E {\bf 70}(6), 066117 (2004).
\bibitem{Caldarelli2002} G. Caldarelli, A. Capocci, P. De Los Rios, M. A. Mu\~noz. Phys. Rev. Lett. {\bf 89}(25), 258702 (2002).
\bibitem{Garlaschelli2004} D. Garlaschelli, M. I. Loffredo. Phys. Rev. Lett. {\bf 93}(18), 188701 (2004).
\bibitem{Mastrandrea2014} R. Mastrandrea, T. Squartini, G. Fagiolo, D. Garlaschelli. New J. Phys. {\bf 16} 043022 (2014).
\bibitem{Gleditsch2002} K. S. Gleditsch. J. Confl. Res. {\bf 46}(5), 712-724 (2002).
\bibitem{foot1a} We use trade volume data for year 2000, expressed in units of $10^8\$$. 
Original volumes were divided by 10 in order to keep the $\{b_i\}_{i=1}^N$ in eq.~(\ref{eq:ecm}) away from 1 and thus avoid numerical instability. 
\bibitem{DeMasi2006} G. De Masi, G. Iori, G. Caldarelli. Phys. Rev. E {\bf 74}(6), 066112 (2006).
\bibitem{foot1b} We consider snapshots of loans aggregated on annual scale (as also done in other works \cite{DeMasi2006}) because of the high volatility of the links at shorter time scales. 
Here we report results for snapshots taken in 1999---yet, analysis for other annual snapshots brings to comparable results. 
\bibitem{Dorogovtsev2010} S. Dorogovtsev. Phys. J. {\bf 9}(11), 51 (2010).
\bibitem{Squartini2011} T. Squartini, D. Garlaschelli. New J. Phys. {\bf 13}, 083001 (2011).
\bibitem{Garlaschelli2009} D. Garlaschelli, M. I. Loffredo. Phys. Rev. Lett. {\bf 102}(3), 038701 (2009).
\bibitem{Garlaschelli2005} D. Garlaschelli, S. Battiston, M. Castri, V. Servedio, G. Caldarelli. Phys. A {\bf 350}(2), 491-499 (2005).
\bibitem{foot3} Comparing $\forall i$ node strength $s_i^*(G_0)$ and $\langle s_i\rangle_{\Omega_{ECM}}=\sum_{j(\neq i)} \langle w_{ij}\rangle$ 
is instead only a consistency check which returns, as it should, an identity (apart from small numerical errors).

\end{thebibliography}
\end{document}